\documentclass[twocolumn]{aastex63}

\shorttitle{Pan-STARRS morphology catalog}
\shortauthors{Goddard \& Shamir}
\graphicspath{{./}{figures/}}

\begin{document}

\title{A catalog of broad morphology of Pan-STARRS galaxies based on deep learning}

\author{Hunter Goddard}
\affiliation{Kansas State University}

\author{Lior Shamir}
\affiliation{Kansas State University}
\email{lshamir@mtu.edu}



\begin{abstract}
Autonomous digital sky surveys such as Pan-STARRS have the ability to image a very large number of galactic and extra-galactic objects, and the large and complex nature of the image data reinforces the use of automation. Here we describe the design and implementation of a data analysis process for automatic broad morphology annotation of galaxies, and applied it to the data of Pan-STARRS DR1. The process is based on filters followed by a two-step convolutional neural network (CNN) classification. Training samples are generated by using an augmented and balanced set of manually classified galaxies. Results are evaluated for accuracy by comparison to the annotation of Pan-STARRS included in a previous broad morphology catalog of SDSS galaxies. Our analysis shows that a CNN combined with several filters is an effective approach for annotating the galaxies and removing unclean images. The catalog contains morphology labels for 1,662,190 galaxies with $\sim$95\% accuracy. The accuracy can be further improved by selecting labels above certain confidence thresholds. The catalog is publicly available. 

\end{abstract}


\section{Introduction} 
\label{sec:intro}

With their ability to generate very large databases, autonomous digital sky surveys have been enabling research tasks that were not possible in the pre-information era, and have been becoming increasingly pivotal in astronomy. The ability of digital sky surveys to image large parts of the sky, combined with the concept of virtual observatories that make these data publicly accessible \citep{djorgovski2001exploration}, has been introducing a new form of astronomy research, and that trend is bound to continue \citep{borne2013virtual,djorgovski2013sky}.

The Panoramic Survey Telescope and Rapid Response System (Pan-STARRS) \citep{kaiser2004pan,flewelling2016pan} is a comprehensive digital sky survey covering $\sim10^3$ degree$^2$ per night using an array of two 1.8m telescopes. Among other celestial objects, Pan-STARRS images a very large number of galaxies. Due to the complexity of galaxy morphology, the ability of current photometric pipelines to analyze these galaxy images is limited, and substantial information that is visible to the humans eye is practically unavailable to users of digital sky surveys data.

To automate the analysis of galaxy images, several methods have been proposed, including GALFIT \citep{pen02}, GIM2D \citep{sim99}, CAS \citep{con03}, the Gini coefficient of the light distribution \citep{Abr03}, Ganalyzer \citep{sha11}, and SpArcFiRe \citep{davis2014sparcfire}. However, the ability of these methods to analyze a large number of real-world galaxy images and produce clean data products is limited, and catalogs of galaxy morphology were prepared manually by professional astronomers \citep{nair2010catalog,efigi}. 

Due to the high volumes of data, the available pool of professional astronomers is not able to provide the sufficient labor to analyze databases generated by modern digital sky surveys, leading to the use of crowdsourcing for that task \citep{lintott2008galaxy,galaxyzoo1,galaxyzoo2}. The main crowdsourcing campaign for analysis of galaxy morphology was Galaxy Zoo \citep{galaxyzoo1}, providing annotations of the broad morphology of galaxies imaged by Sloan Digital Sky Survey (SDSS), as well as other surveys such as the Cosmic Assembly Near-infrared Deep Extragalactic Legacy (CANDELS). However, analyzing the broad morphology of SDSS galaxies required $\sim$3 years of work performed by over $10^5$ volunteers, and led to $\sim7\cdot10^4$ galaxies considered ``superclean". Given the huge databases of current and future sky surveys, it is clear that even when using crowdsourcing, the throughout of manual classification might not be sufficient for an exhaustive analysis of such databases.

The use of machine learning provided more effective methods for the purpose of galaxy image classification  \citep{sha09,huertas2009robust,banerji2010,shamir2013automatic,schutter2015galaxy,kum14,dieleman2015rotation,hocking2017automatic,kuminski2018hybrid,silva2018sparcfire}, and the use of such methods also provided computer-generated catalogs of galaxy morphology \citep{Hue10,sim11,shamir2014automatic,kum16,huertas2015catalog,huertas2015morphologies,timmis2017catalog,paul2018catalog,shamir2019automatic}. Automatic annotation methods were also tested on Pan-STARRS data by using the photometric measurements of colors and moments, classified by a Random Forest classifier to achieve a considerable accuracy of $\sim89$\% \citep{baldeschi2020star}. 

Here we use automatic image analysis to prepare a catalog of the broad morphology of $\sim1.7\cdot10^6$ Pan-STARRS DR1 galaxies. The catalog was generated by using a data analysis process that involves several steps and two convolutional neural networks (CNN) that automated the annotation process to handle the high volume of data.

\section{Data}
\label{sec:data}

The galaxy image data is sourced from the first data release (DR1) of Pan-STARRS \citep{hodapp2004design,flewelling2016pan,chambers2016pan}. First, all objects with Kron r magnitude of less than than 19 and identified by Pan-STARRS photometric pipeline as extended in all bands were selected. 

\begin{figure}[hpt]
    \centering
    \includegraphics[scale=0.7]{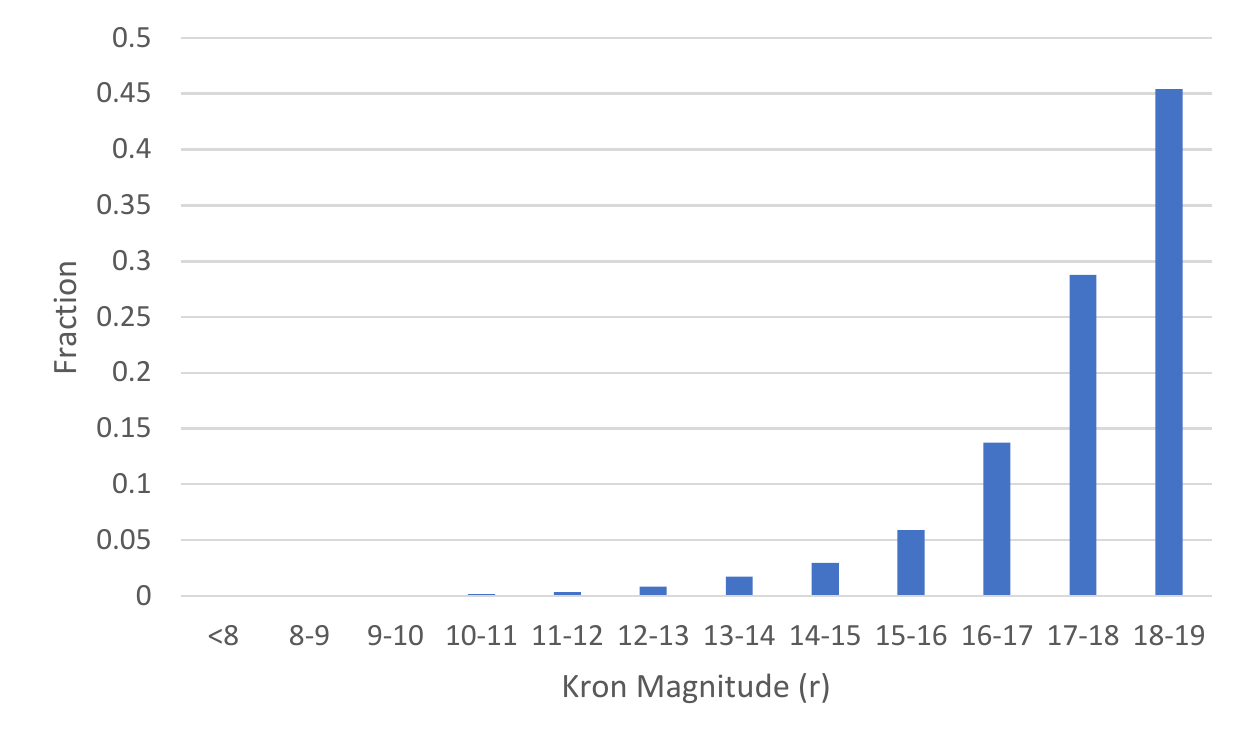}
    \caption{\textit{Distribution of the r Kron magnitude of the galaxies in the dataset.}}
    \label{fig:magnitude}
\end{figure}

To filter objects that are too small to identify morphology, objects that have Petrosian radius smaller than 5.5'' were removed. To remove stars, objects that their PSF i magnitude subtracted by their Kron i magnitude was greater than 0.05 were also removed. That led to a dataset of 2,394,452 objects \citep{timmis2017catalog}. Objects that were flagged by Pan-STARRS photometric pipeline as artifacts, had a brighter neighbor, defect, double PSF, or a blend in any of the bands were excluded from the dataset. That led to a dataset of 2,131,371 objects assumed to be sufficiently large and clean to allow morphological analysis. Figure~\ref{fig:magnitude} shows the distribution of the r Kron magnitude of the galaxies in the dataset.

The galaxy images were then downloaded using Pan-STARRS {\it cutout} service. The images are in the JPG format and have a dimensionality of 120$\times$120 pixels as in \citep{kum16}. Pan-STARRS {\it cutout} provides JPG images for each of the bands. Here we use the images of the g band, as the color images using the y, i, and g bands are in many cases noisy, and do not allow effective analysis of the morphology. The process of downloading the data was completed in 62 days.

The initial scale of the {\it cutout} was set to 0.25'' per pixel. For each image that was downloaded, all bright pixels (grayscale value higher than 125) located on the edge of the frame were counted. If more than 25\% of the pixels on the edge of the frame were bright, it is an indication that the object does not fully fit inside the frame. In that case, the scale was increased by 0.05'', and the image was downloaded again. That was repeated until the number of bright pixels on the edge was less than 25\% of the total edge pixels, meaning that the object is inside the frame. The JPG images are far smaller than the FITS images. A 120$\times$120 JPG image retrieved through Pan-STARRS {\it cutout} service is normally of size of $\sim$3KB, while an image of the same dimensions in the FITS format will be $\sim$76KB. Although the FITS files provide more information, downloading the files in FITS format is substantially slower. While downloading the JPG images lasted 62 days, downloading the same number of images in the much larger FITS format will require a far longer period of time. The JPG images do not allow photometry, but they are smaller than the FITS files and provide visual information about the shape of the galaxy, which is the information required for the morphological classification of the galaxies. As explained in Section~\ref{subsec:classify1}, the training of the neural network was done with images retrieved from Pan-STARRS, with the exact same size and format as the images that were annotated by the neural network after it was trained.

\section{Image analysis method} 
\label{sec:method}

The filtering of the data described in Section~\ref{sec:data} aims at removing objects that are not clean galaxy images. That allows to reduce the number of images downloaded and classified in the next step with the deep neural network. The removal of objects that are not galaxy images also makes the neural network more accurate due to the higher consistency of the data it is trained with.

To remove saturated images and images that have too few features to allow morphological classification, two additional filters are used. The first filter finds the ratio of fully saturated pixels (a grayscale value of 255 in the JPG image) to the total number of pixels and discards the image if this ratio is higher than 15:1000. Since a high number of saturated pixels is not expected in a clean galaxy image, the simple threshold of 1.5\% is sufficient to identify and reject saturated images that are not galaxy images. This step rejected 30,220 objects that were identified as saturated.

The second filter uses the Otsu global threshold method \citep{otsu1979threshold} to separate the image into foreground and background pixels. If the number of foreground pixels is less than 1.8\% of the total image, the image is marked as having too few distinguishable features. This filter rejected 375,107 galaxies that were identified as having too little foreground to allow identification. Together, these filters removed 405,327 images ($\sim$19\%) from the data set. The thresholds were determined experimentally by observing galaxy image samples. Table~\ref{filtered_images} shows examples of several objects that were filtered based on too few foreground pixels or too many saturated pixels.

\begin{table}[htbp]
\caption{Examples of images filtered for having too few foreground pixels or having too many saturated pixels.}
\begin{center}
\begin{tabular}{|c|c|c|}
\hline
Image & Saturated  & Foreground  \\
         & pixels (\%)  & pixels (\%)  \\
\hline
\includegraphics[scale=0.5]{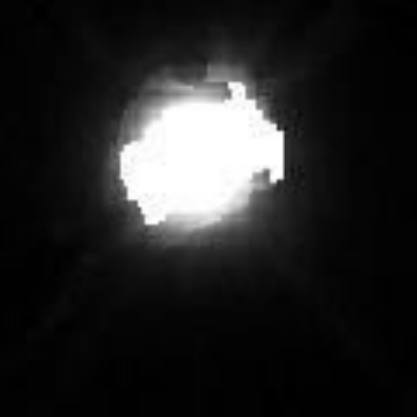} & 6.1 & 10.7 \\
\hline
\includegraphics[scale=0.5]{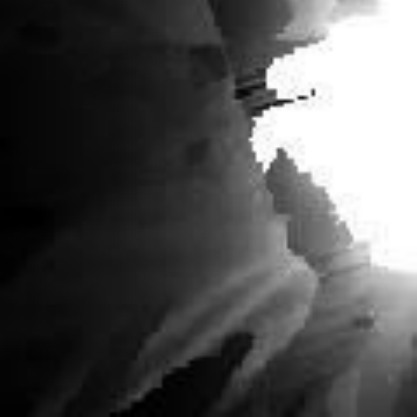} & 13.5 & 21.7 \\
\hline
\includegraphics[scale=0.5]{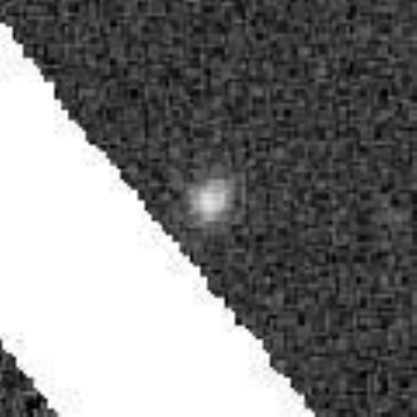} & 30.9 & 34.9 \\
\hline
\includegraphics[scale=0.5]{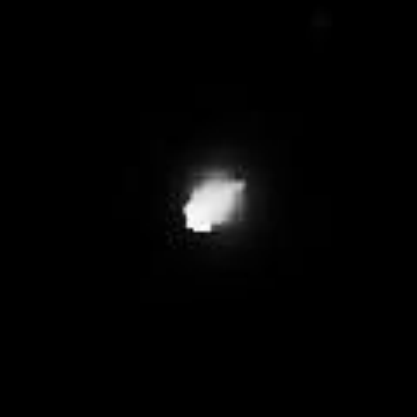} & 0.06 & 1.4 \\
\hline
\includegraphics[scale=0.5]{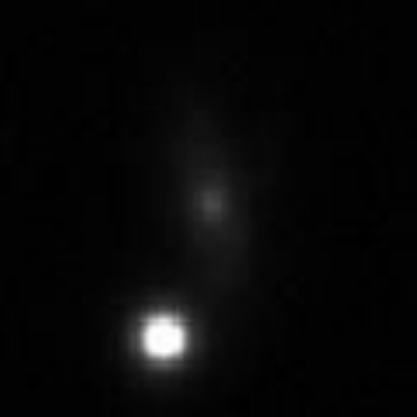} & 0.16 & 1.1 \\
\hline
\end{tabular}
\label{filtered_images}
\end{center}
\end{table}

\subsection{Primary classification}
\label{subsec:classify1}

The classifier used for the purpose of annotating the galaxy images is a deep convolutional neural network (DCNN) based on the LeNet-5 architecture \citep{lecun1998gradient}. To adjust the model for input images of size 120$\times$120 instead of 32$\times$32, the kernel in the first convolutional layer was changed from 5$\times$5 with stride 1 to 10$\times$10 with stride 2, and the filter in the first pooling layer was similarly changed from 2$\times$2 with stride 2 to 4$\times$4 with stride 4. Each of the following layers has identical hyperparameters except for the output layer, where the number of classes is reduced from 10 to 2. The SoftMax output layer of the model provides a degree of certainty for the annotations that allows controlling the size/accuracy trade-off of the catalog, as will be discussed in Section~\ref{sec:results}.

Training samples were obtained using the debiased ``superclean" Galaxy Zoo annotations. ``Superclean" objects are objects on which 95\% or more of the annotators agreed on their morphology with correction for the redshift bias \citep{galaxyzoo1}. That selection leads to a subset of very consistent annotations \citep{galaxyzoo1}, but it also filters the vast majority of Galaxy Zoo annotations that do not satisfy these requirements. The Galaxy Zoo crowdsourcing campaign annotated galaxies imaged by SDSS, which is a different instrument with a different image processing pipeline. Although it has been shown in the past that neural networks trained with data from one telescope can be used to classify data acquired by other telescopes \citep{dominguez2019transfer}, it has also been shown that the accuracy of such networks is inferior to the accuracy of a neural network trained and tested with data from the same instrument \citep{dominguez2019transfer}. Since a large number of galaxies annotated by Galaxy Zoo were also imaged by Pan-STARRS, the Pan-STARRS images of these galaxies can be fetched and be used as the training data, so that the images used to train the neural network are imaged by the same instrument that imaged the galaxies annotated by that network. 

Due to the substantial overlap between the footprint of Pan-STARRS and SDSS, the idea of using SDSS data as labels to train machine learning systems with Pan-STARRS data has been used in the past. For instance, \cite{tarrio2020photometric} used spectroscopic data from SDSS as labels for training a machine learning system that can determine the photometric redshift of Pan-STARRS galaxies.

In order to train the neural network with images from the same instrument that it is expected to annotate, the images of the galaxies annotated by Galaxy Zoo were retrieved from Pan-STARRS. Pan-STARRS has a different footprint than SDSS, so not all galaxies annotated by Galaxy Zoo are also imaged by Pan-STARRS. However, 22,456 Galaxy Zoo galaxies with ``superclean'' annotations were matched with galaxies in Pan-STARRS DR1 based on their right ascension and declination (within difference of 0.0001 degrees). These images were fetched from Pan-STARRS and were used for training the neural network.

\begin{figure}[hpt]
    \centering
    \includegraphics[scale=0.7]{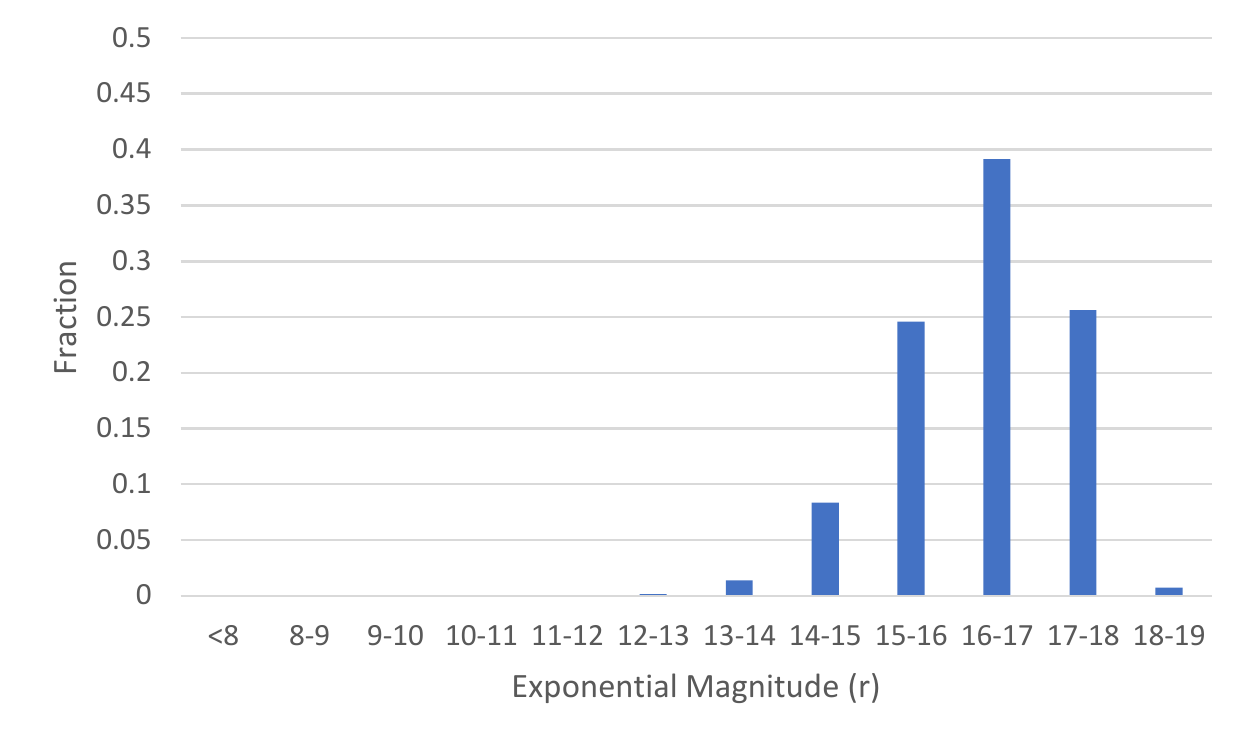}
    \caption{\textit{Distribution of the r exponential magnitude of the galaxies in the Galaxy Zoo dataset from which the annotations were taken.}}
    \label{sdss_magnitude}
\end{figure}

Figure~\ref{sdss_magnitude} shows the distribution of the r exponential magnitude of the Galaxy Zoo galaxies that their annotations were used for the compilation of the training set. The magnitude distribution is somewhat different from the magnitude distribution of the Pan-STARRS galaxies shown in Figure~\ref{fig:magnitude}, which can be explained by the 17.77 limiting r Petrosian magnitude applied to the initial Galaxy Zoo sample \citep{lintott2008galaxy}. As mentioned above, the SDSS images themselves were not used for the training.

\begin{figure}[hpt]
    \centering
    \includegraphics[scale=0.5]{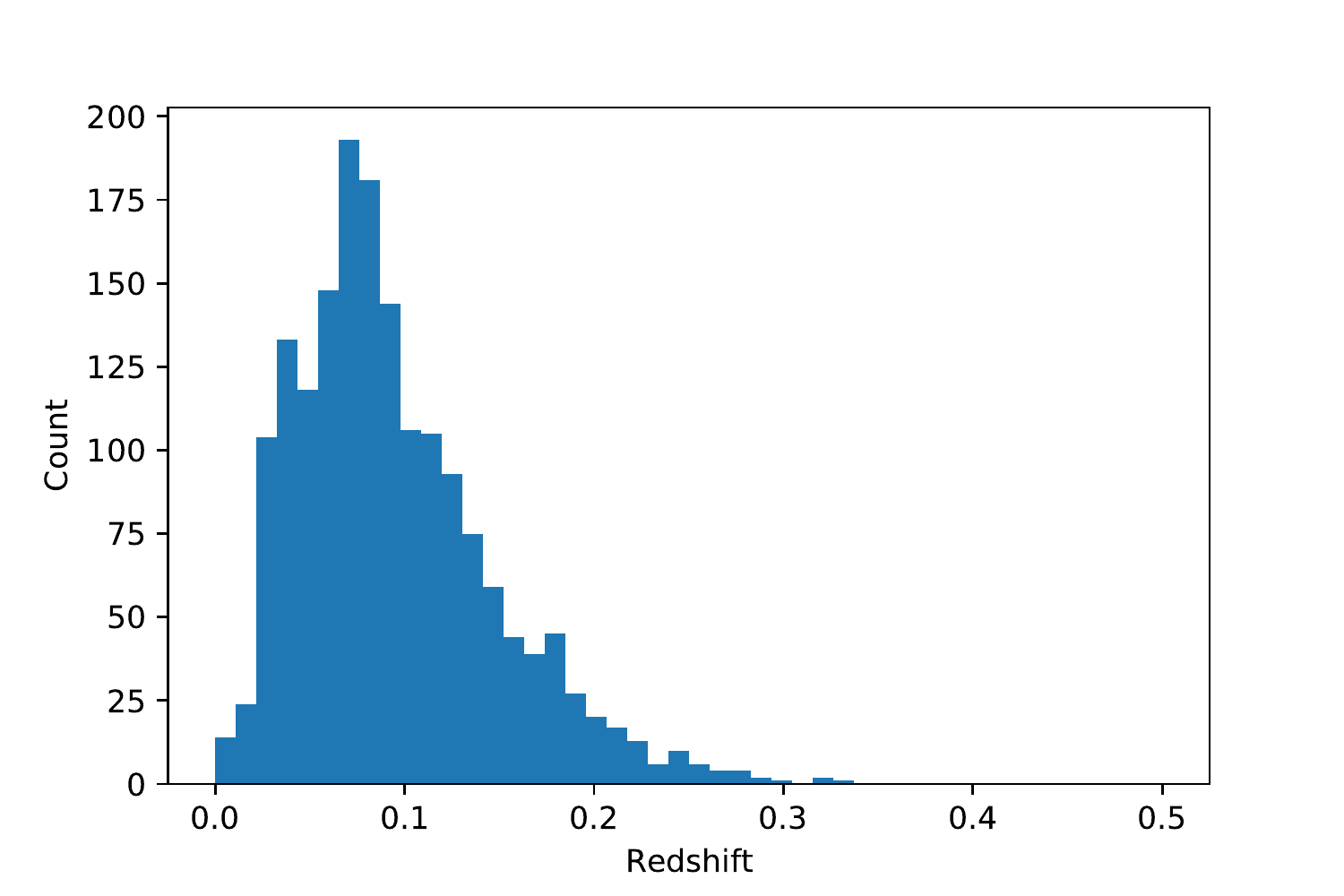}
    \caption{\textit{Distribution of the redshift of the galaxies in the Pan-STARRS dataset. The redshift values were taken from SDSS.}}
    \label{z_hist_ps}
\end{figure}

\begin{figure}[hpt]
    \centering
    \includegraphics[scale=0.5]{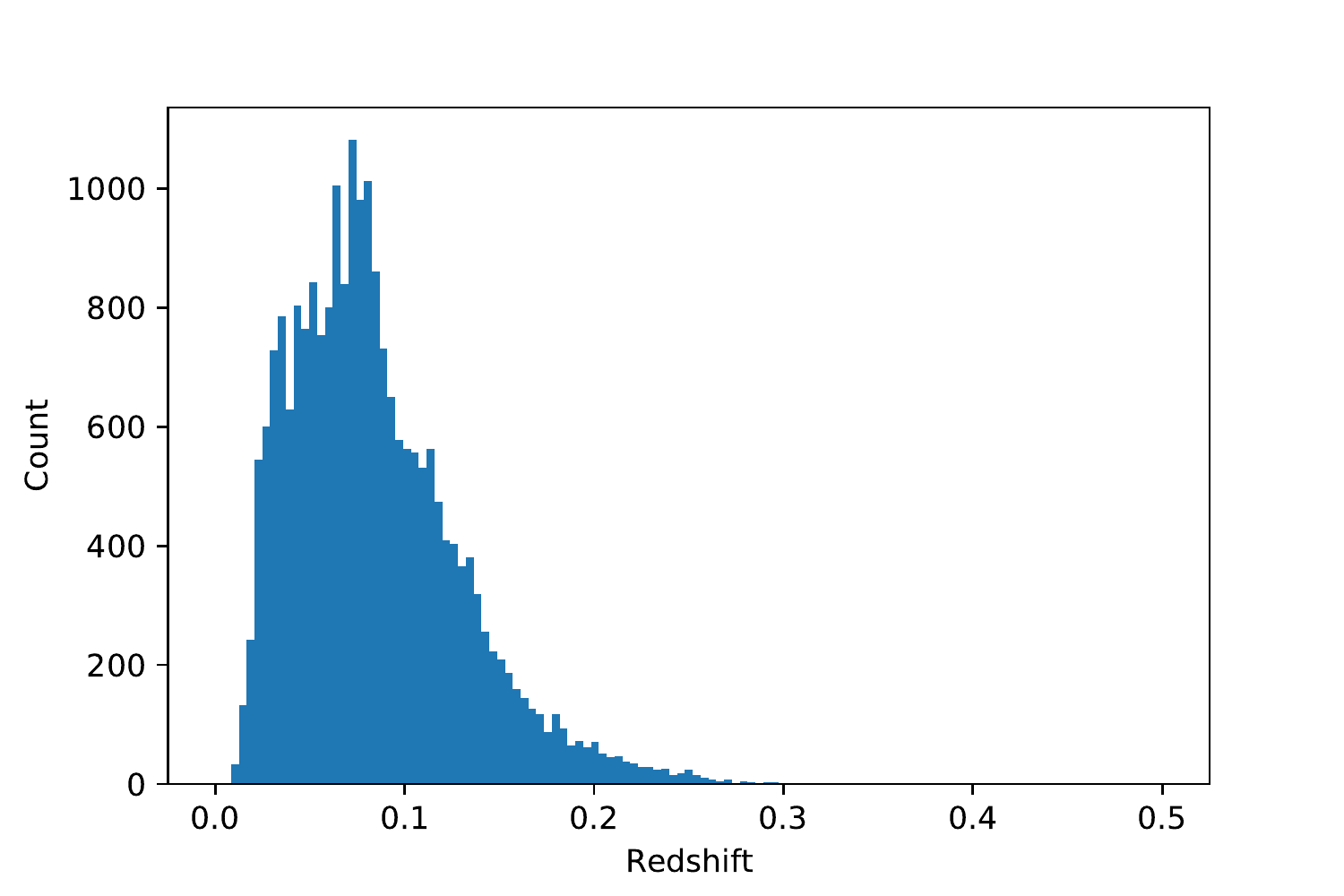}
    \caption{\textit{Distribution of the redshift of the galaxies in the Galaxy Zoo dataset from which the annotations were taken.}}
    \label{z_hist_gz}
\end{figure}

Figures~\ref{z_hist_ps} and~\ref{z_hist_gz} show the histograms of the redshift distribution of the galaxies in Pan-STARRS and SDSS, respectively. The number of Pan-STARRS galaxies with redshift is small since Pan-STARRS does not collect spectra, and the spectra was only taken from SDSS galaxies that overlapped with Pan-STARRS galaxies. The two graphs show that the distribution of the redshift is similar in both datasets. 

Galaxy Zoo manual annotations have been shown in the past to be sensitive to the spin direction of the galaxies \citep{land2008galaxy}. To eliminate the possible effect of spin patterns, the training set was augmented such that all galaxies were mirrored, and both the original and mirrored image of each galaxy were used in the training set. That resulted in a training set of 31,564 spiral images and 13,348 elliptical images. Mirroring the spiral galaxies ensures a symmetric dataset that is not biased by certain preferences of the human volunteers who annotated the galaxies. That is, while mirroring the images in the training set is often used when training deep neural networks for augmenting the data and increasing the number of training samples, in this case it was also used to produce a symmetric unbiased dataset. Mirroring of the elliptical galaxies was done to ensure consistency in the manner training data are handled, and avoid a situation in which different classes are handled differently. 

\begin{figure}[hpt]
    \centering
    \includegraphics[scale=0.3]{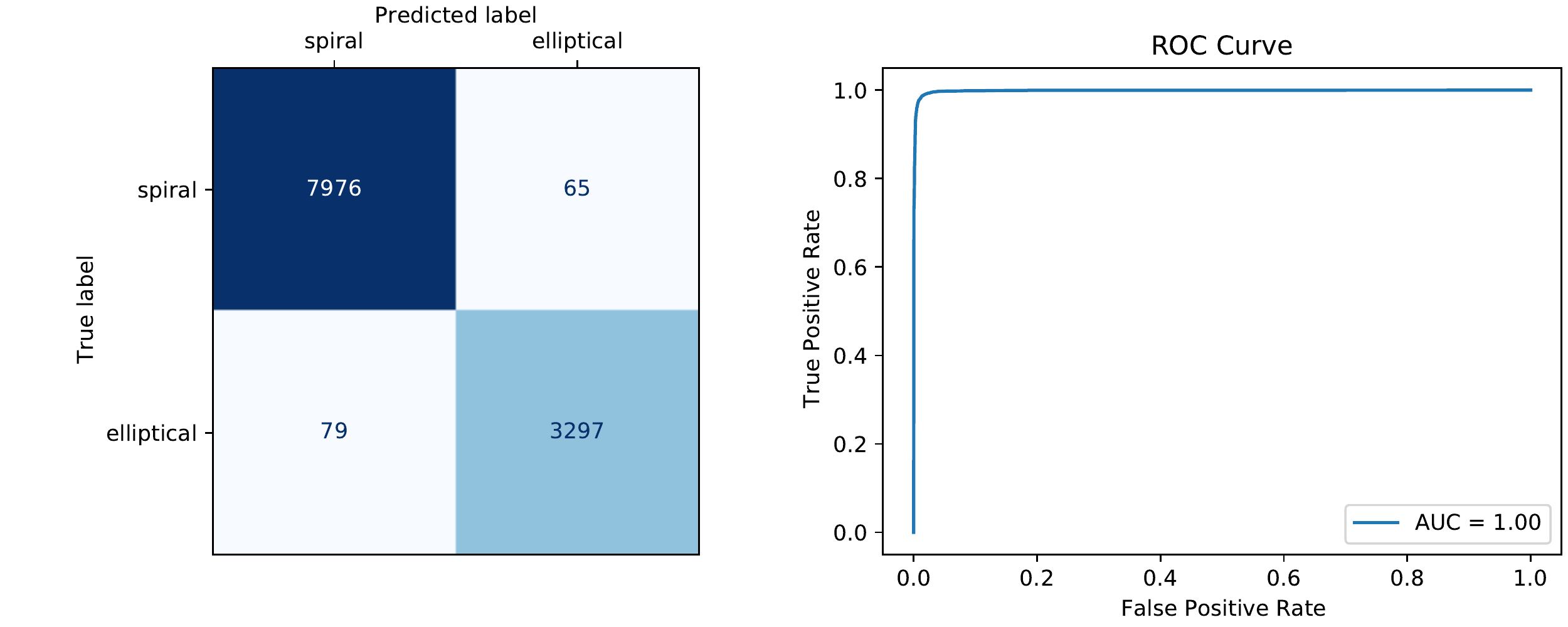}
    \caption{Confusion matrix and ROC curve of the classification of the 30\% test samples.}
    \label{cm_roc_training}
\end{figure}

The classifier is implemented in Python 3 using TensorFlow \citep{tensorflow2015-whitepaper} and Keras \citep{chollet2015keras}. The model was trained for 250 epochs on a 70\% training subset and ended with 98.7\% accuracy when evaluated against the remaining 30\% testing subset. Figure~\ref{cm_roc_training} shows the confusion matrix and receiver operating characteristic (ROC) curve of the classification. The high accuracy shows that although the galaxy images were labeled with annotations made with SDSS galaxies, the annotations were still consistent in Pan-STARRS images. That consistency indicates that the two sky surveys are roughly equivalent in the information they provide about the morphology of the galaxies.

Loss was computed using categorical cross entropy, and stochastic gradient descent (SGD) was used as the optimizer. Various activation functions including ReLU were tested, and gave comparable classification accuracy. The $\tanh$ activation used by LeNet-5 gave the best performance and therefore was used for the model. Classification on the total data set (excluding those removed by the filtering step) labeled 904,550 images as elliptical galaxies and 821,494 images as spiral galaxies.

\subsection{Secondary Classification}
\label{subsec:classify2}

Following the classification described in Section~\ref{subsec:classify1}, the set of images predicted as spiral was shown to contain a significant number of ``ghosts", or unclean images. The CNN classifier interpreted the unclean images as patterns of spiral features, leaving the elliptical predictions relatively clean.

To remove these ghosts, we constructed a second deep CNN to separate them from the true spirals. The architecture of this model is simpler than the first, using three convolutional layers with filter sizes of 7$\times$7$\times$8, 5$\times$5$\times$32, and 3$\times$3$\times$64, ReLU activation function, and a single SoftMax output layer. Between the convolutional layers there are max pooling layers that each reduce the input dimensions by half. The model uses the Adam optimizer and categorical cross entropy for loss.

\begin{figure}[hpt]
    \centering
    \includegraphics[scale=0.3]{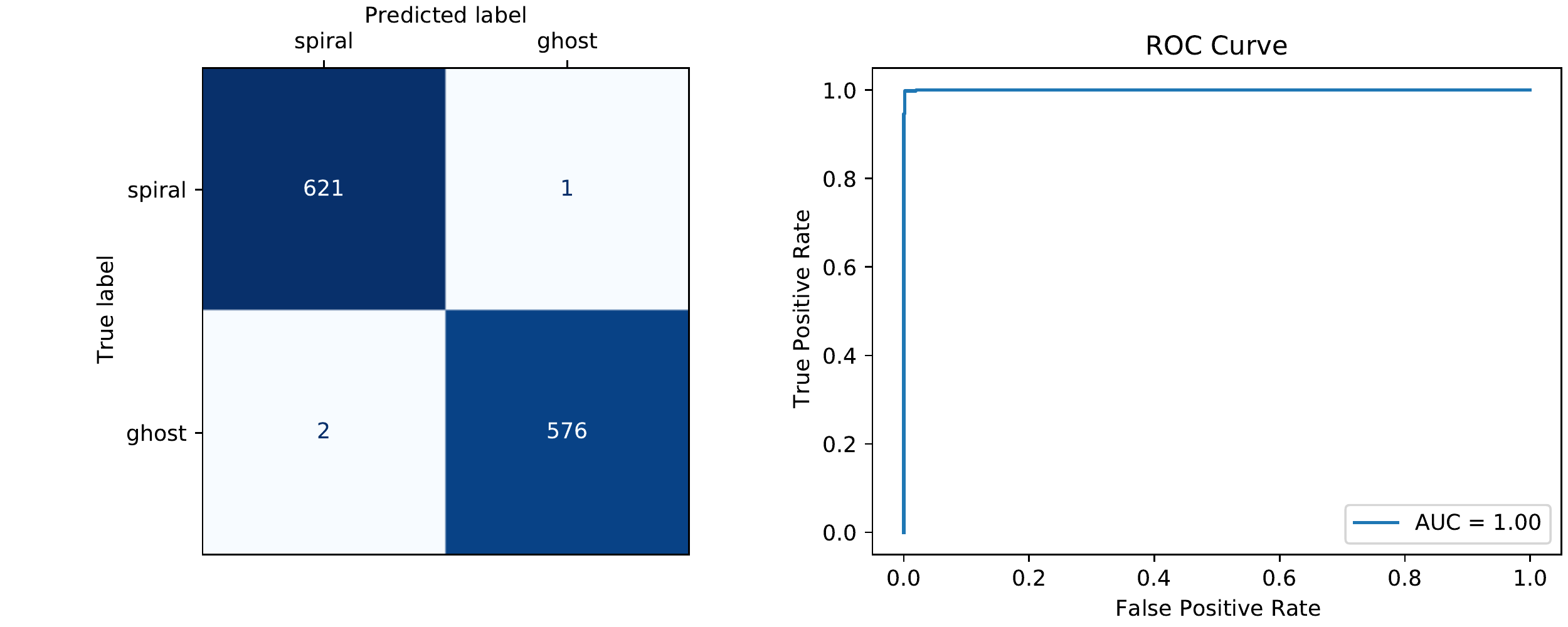}
    \caption{Confusion matrix and ROC curve of the classification of the 1,200 ghosts and spiral galaxies.}
    \label{ghost_cm_roc}
\end{figure}

For training, several hundred ghost images were initially selected from the set of galaxy images that were mistakenly predicted as spirals, and an equal number of spiral galaxy images were randomly selected from the original spiral training set. These images were divided into 70\% training and 30\% testing subsets as before. The model converged during training, and the images originally labeled as spirals were further classified into true spirals and ghosts. This process was repeated several times by selecting additional training images from those labeled as ``ghosts" until the size of the training set reached 4,000 images. Testing the neural network shows that the network identifies ``ghosts'' with accuracy very close to 100\%, and almost no false positives. Figure~\ref{ghost_cm_roc} shows the confusion and matrix and ROC curve when testing 1,200 images of spiral galaxies and ghosts. The final iteration of this classifier identified a total of 63,854 images as ``ghosts" ($\sim7.8\%$), removing them from the set of spiral galaxies.

\section{Results}
\label{sec:results}

\begin{figure}[hpt]
    \centering
    \includegraphics[scale=0.75]{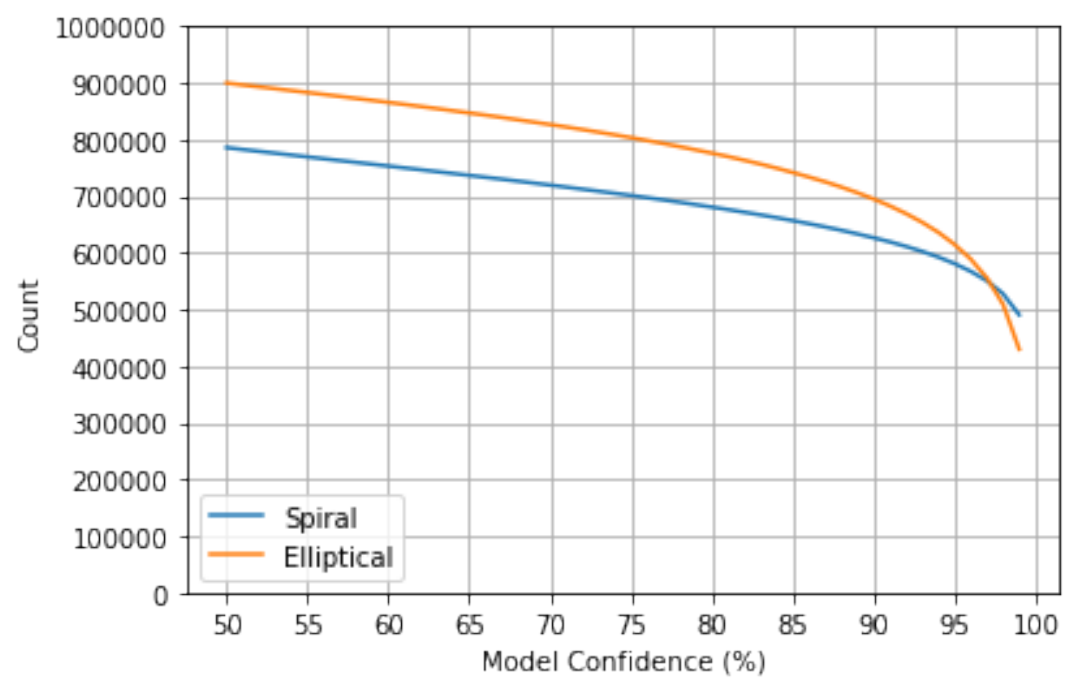}
    \caption{Number of spiral and elliptical galaxies remaining when keeping only those at or above a certain model confidence.}
    \label{fig:counts}
\end{figure}

The application of the methods described in Section~\ref{sec:method} to the Pan-STARRS images described in Section~\ref{sec:data} provided a catalog of 1,662,190 galaxies. The catalog is accessible through a simple CSV file that can be downloaded at \url{https://figshare.com/articles/PanSTARRS_DR1_Broad_Morphology_Catalog/12081144}. Each row in the catalog is a galaxy, and includes the Pan-STARRS object ID of the galaxy, its right ascension, declination, and the probability of the galaxy to be spiral or elliptical as estimated by the SoftMax layer of the CNN as described in Section~\ref{sec:method}. Figure~\ref{fig:counts} shows the number of galaxies available after applying a threshold to the output of the SoftMax layer of the model.

The catalog includes 904,550 galaxies identified as elliptical and 757,640 identified as spiral. It should be noted that the annotation of a galaxy as an elliptical galaxy means that no spiral features were identified. However, the ability of an algorithm or a person to identify spiral features largely depends on the ability of the optics to provide a detailed image. Therefore, the identification of a galaxy as elliptical does not necessarily guarantee that the galaxy indeed does not have spiral features, but that the optics cannot identify such features \citep{dojcsak2014quantitative}. For instance, Table~\ref{hubble} shows examples of galaxies imaged by Pan-STARRS and SDSS, and the same galaxies imaged by Hubble Space Telescope (HST). As the table shows, these galaxies do not have clear visible spiral arms in the Earth-based telescopes, while the arms are seen clearly in the HST images. 

\begin{table*}[ht]
\caption{Galaxies imaged by Pan-STARRS, SDSS, and HST. While the Pan-STARRS and SDSS images do not show clear spiral arms of the galaxies, HST shows that these galaxies are clearly spiral, and the arms can be identified.}
\begin{center}
\begin{tabular}{|c|c|c|c|}
\hline
Coordinates & Pan-STARRS & SDSS & HST \\
\hline
(150.165$^o$,1.588$^o$)  & \includegraphics[scale=0.45]{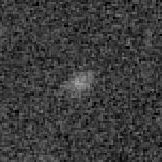} &  \includegraphics[scale=0.75]{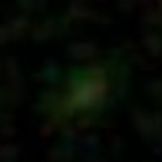}  & \includegraphics[scale=0.45]{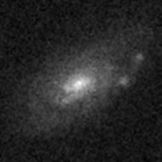}  \\
\hline
(150.329$^o$,1.603$^o$)  & \includegraphics[scale=0.45]{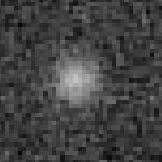}  & \includegraphics[scale=0.75]{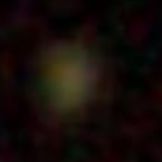} & \includegraphics[scale=0.45]{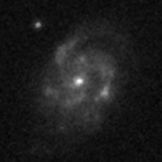} \\
\hline
(149.951$^o$,1.966$^o$) & \includegraphics[scale=0.45]{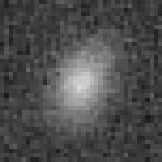} & \includegraphics[scale=0.75]{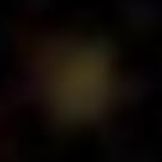} & \includegraphics[scale=0.45]{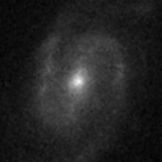} \\
\hline
\end{tabular}
\label{hubble}
\end{center}
\end{table*}

\subsection{Comparison to an existing SDSS catalog}
\label{result_comp}

In the absence of a large manually annotated galaxy morphology catalog of Pan-STARRS galaxies, the evaluation of the consistency of the annotations was done using annotations of SDSS galaxies that were also imaged by Pan-STARRS. The largest catalog of broad morphology of SDSS galaxies is \citep{kum16}, with annotation of $\sim3\cdot10^{6}$ galaxies. Although SDSS is a different sky survey, the footprint of SDSS overlaps with the footprint of Pan-STARRS. Since the \citep{kum16} catalog is large, it is expected that some galaxies in \citep{kum16} will also be included in the catalog of Pan-STARRS galaxies described in this paper.

\begin{figure}
    \centering
    \includegraphics[scale=0.85]{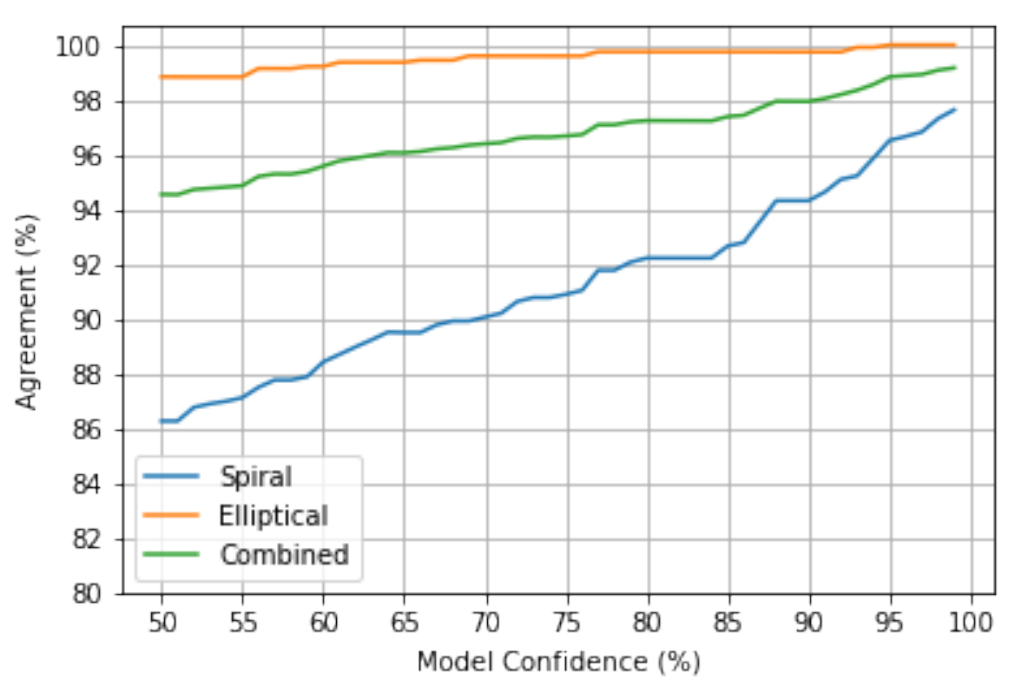}
    \caption{The proportion of predicted labels that, when restricted to a minimum confidence threshold, agree with the annotations in \citep{kum16}. For example, restricting the catalog to labels with 90\% confidence or higher will have approximately 98\% agreement with the annotations in \citep{kum16}}.
    \label{fig:agreement}
\end{figure}

To evaluate the catalog, the annotations were compared to the annotations of SDSS galaxies in \citep{kum16} with high degree of confidence of the annotations. Since the images of \citep{kum16} are collected and processed by the SDSS pipeline, their object identifiers naturally do not match the identifiers of Pan-STARRS objects. Therefore, the objects were matched by their coordinates, with tolerance of 0.0001$^o$ to account for subtle differences in measurements between the two telescopes. This produced 13,186 total matches with 1,961 having 90\% or higher confidence in the \citep{kum16} catalog. Figure~\ref{fig:agreement} shows the degree of agreement between the annotations of the galaxies in the catalog and the annotations of the galaxies in \citep{kum16} with high confidence level.

When comparing the accuracy of the catalog to the accuracy of \citep{kum16}, the algorithm was more accurate in identifying spiral galaxies, while the algorithm used in this catalog was more accurate in the identification of elliptical galaxies. The algorithm used in \citep{kum16} is a ``shallow learning'' algorithm \citep{shamir2008wndchrm}, which is a different paradigm of machine learning compared to the deep convolutional neural network used here. Shallow learning features such as textures and fractals might better reflect spiral arms, and therefore increase the ability of the algorithm to detect spiral galaxies. Elliptical galaxies are more consistent in shape than spiral galaxies, which can increase the performance of deep convolutional neural networks that their accuracy depend on the consistency of the images.   

\begin{figure}
    \centering
    \includegraphics[scale=0.5]{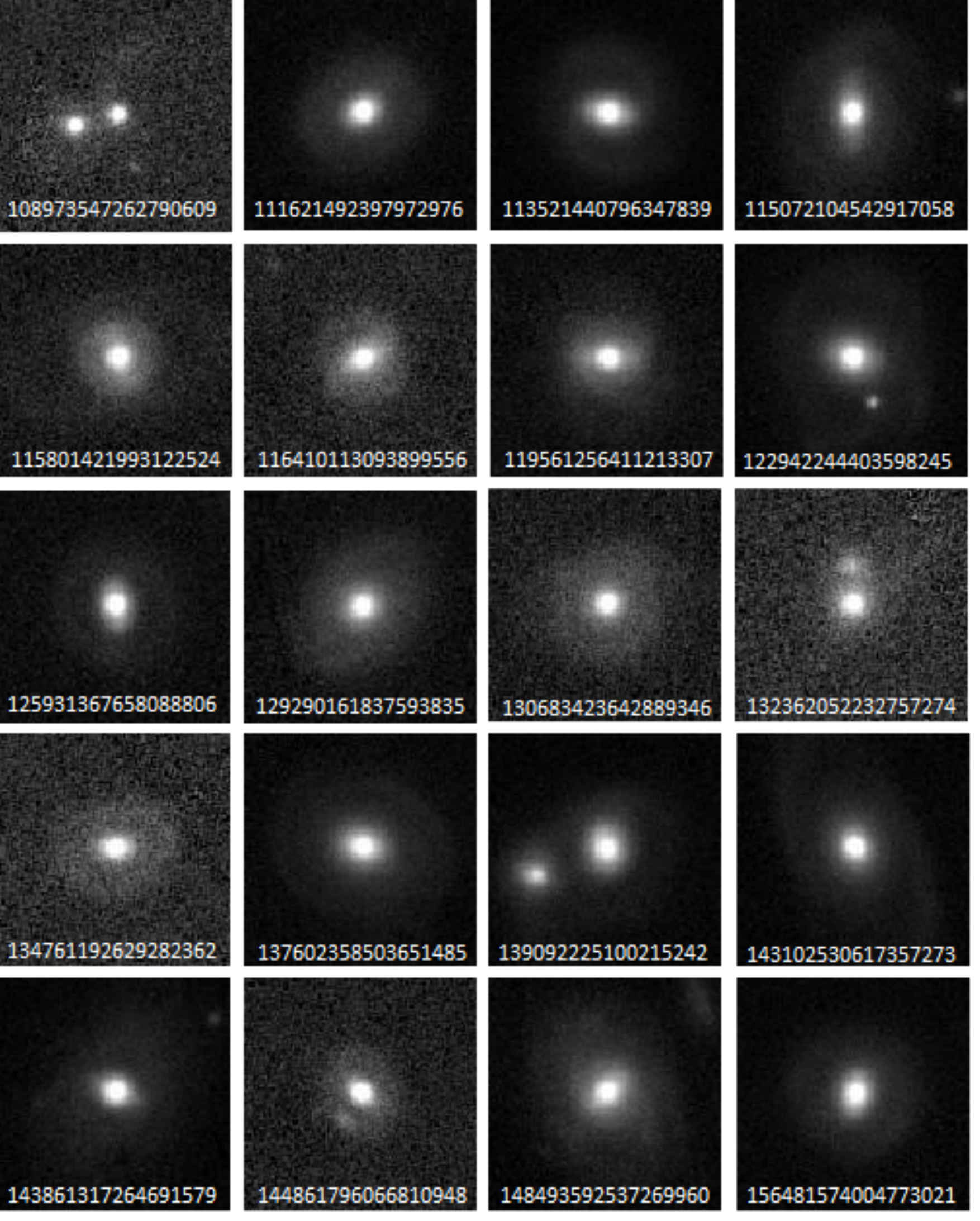}
    \caption{Galaxies imaged by Pan-STARRS that were classified incorrectly as elliptical in \citep{kum16} and as spiral in this catalog.}
    \label{kum_spiral}
\end{figure}

\begin{figure}
    \centering
    \includegraphics[scale=0.5]{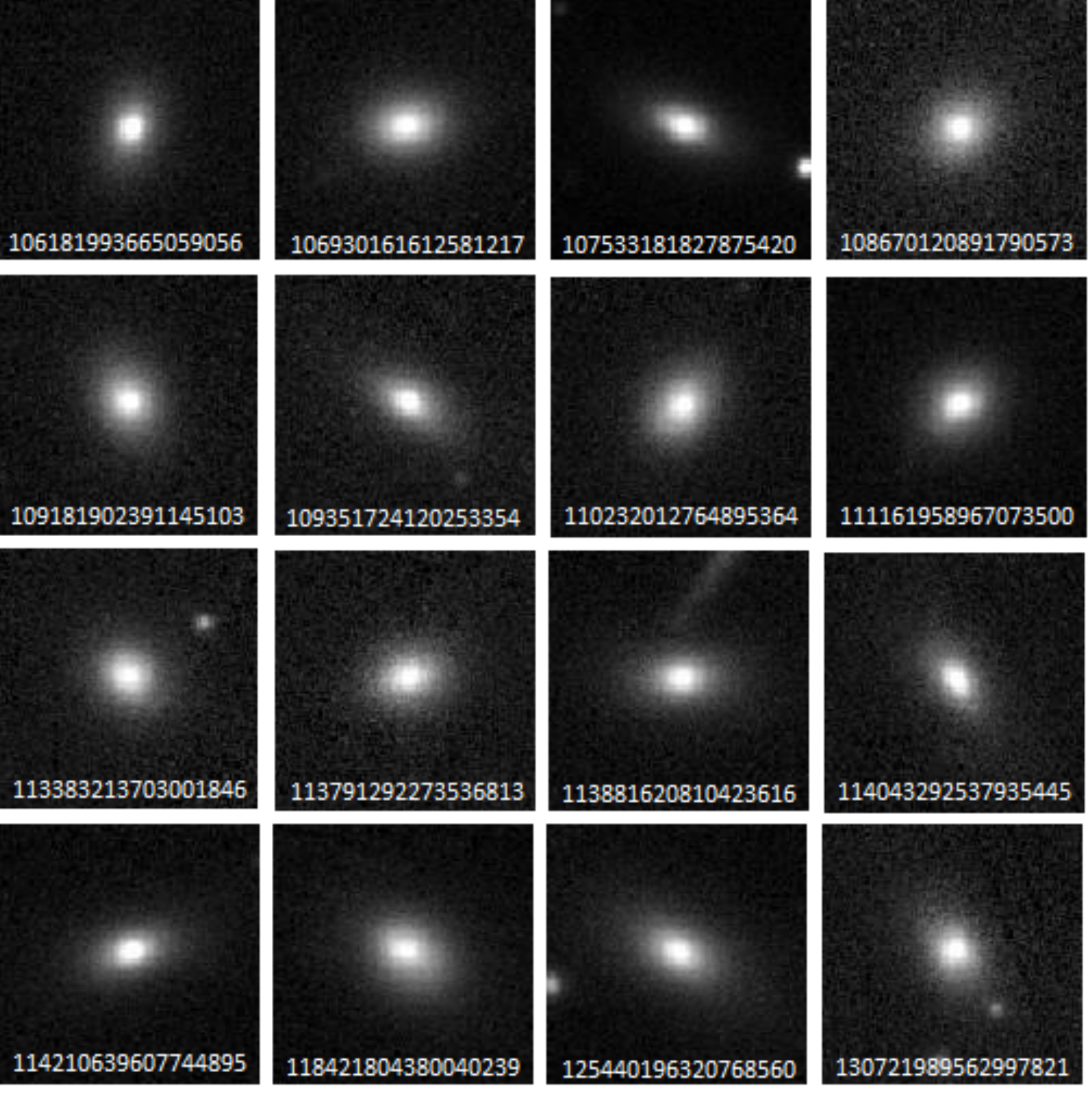}
    \caption{Galaxies imaged by Pan-STARRS that were classified as spiral in \citep{kum16} but as elliptical in this catalog.}
    \label{kum_elliptical}
\end{figure}

Figure~\ref{kum_spiral} shows galaxies that were classified as elliptical galaxies by the \citep{kum16} catalog, but as spiral in this catalog, and by visual inspection seem spiral galaxies. Figure~\ref{kum_elliptical} shows galaxies classified in \citep{kum16} as spiral but in this catalog as elliptical. Careful manual inspection of the images show that the galaxies in Figure~\ref{kum_spiral} are spiral galaxies, but in many of the cases the arms are dim. It is therefore possible that the shallow learning algorithm used in \citep{kum16} failed to detect these spiral galaxies due to the weak presence of the spiral arms. 

Figure~\ref{kum_elliptical} shows that the galaxies classified as elliptical do not have clear spiral arms. However, given that the resolution of Pan-STARRS is limited, it is possible that these galaxies are spiral, as shown in Figure~\ref{hubble}, where spiral arms not visible in Pan-STARRS become clearly visible using a space-based instrument with higher resolution.

\begin{figure}
    \centering
    \includegraphics[scale=0.5]{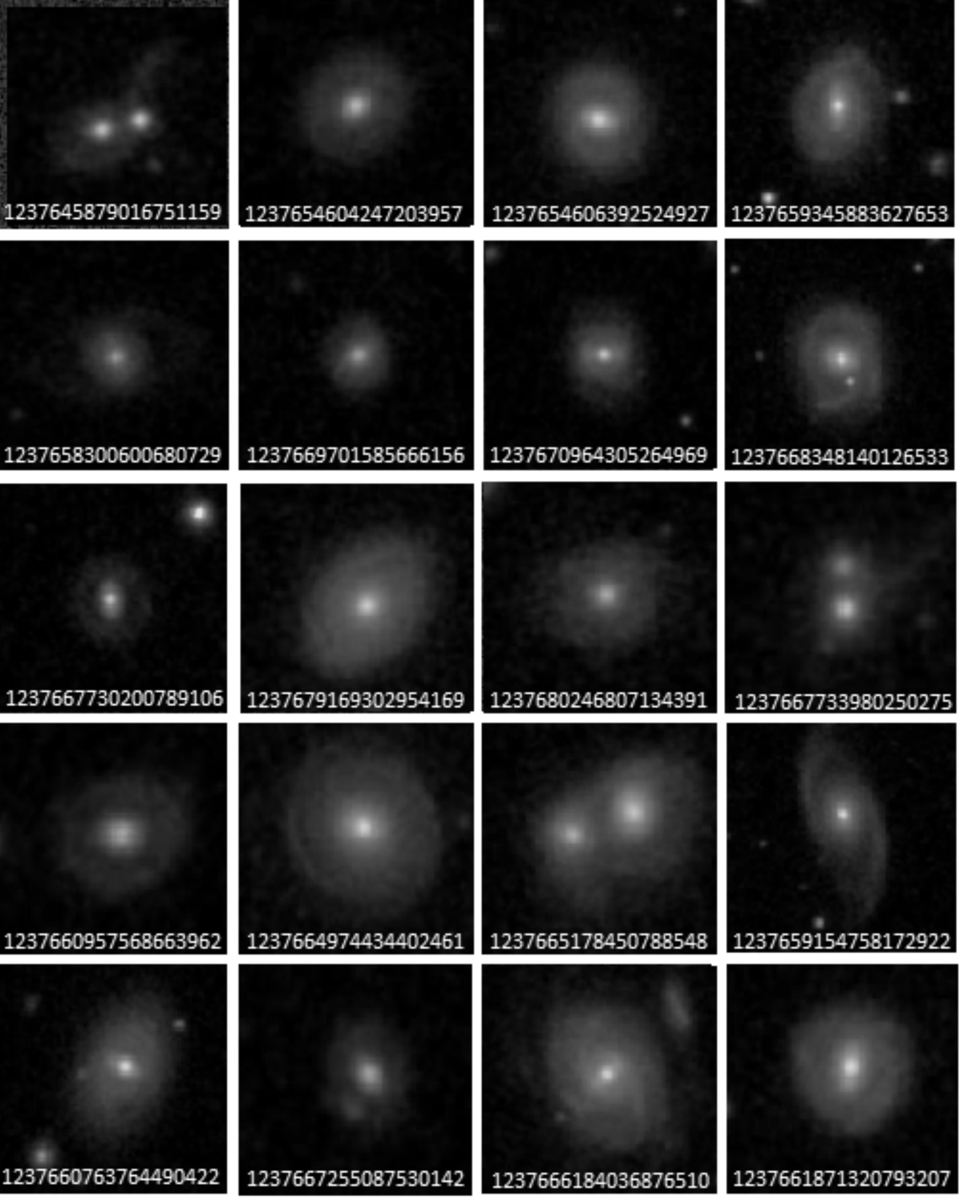}
    \caption{Galaxies imaged by SDSS and were classified incorrectly as elliptical in \citep{kum16}.}
    \label{kum_spiral_SDSS}
\end{figure}

\begin{figure}
    \centering
    \includegraphics[scale=0.5]{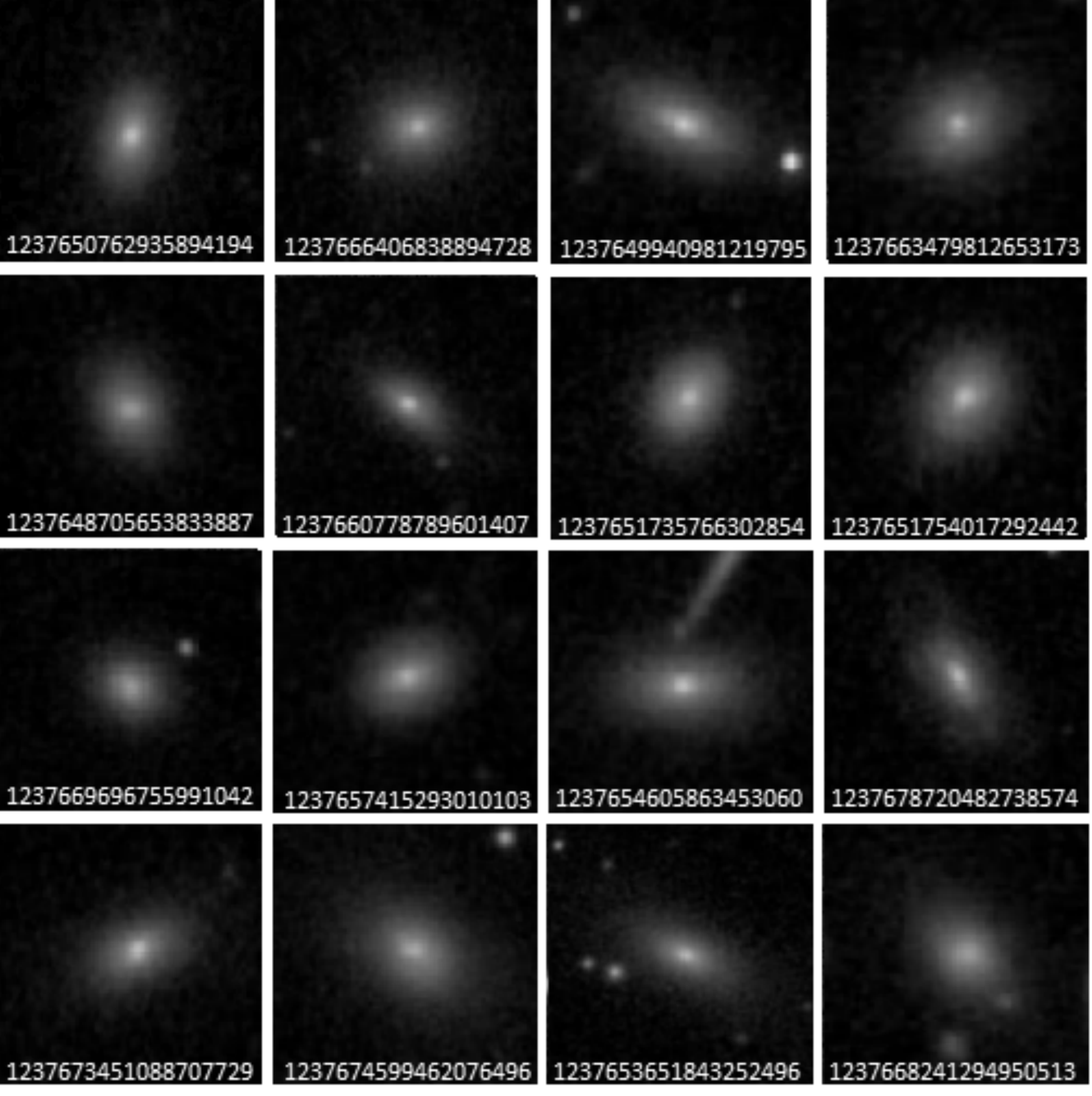}
    \caption{Galaxies imaged by SDSS and were classified as spiral in \citep{kum16} but as elliptical in this catalog.}
    \label{kum_elliptical_SDSS}
\end{figure}

Figures~\ref{kum_spiral_SDSS} and~\ref{kum_elliptical_SDSS} show the same galaxies in Figures~\ref{kum_spiral} and~\ref{kum_elliptical} imaged by SDSS, and classified as elliptical in \citep{kum16}. Figure~\ref{kum_spiral_SDSS} shows that some of the galaxies are ring galaxies or interacting systems, while some of them are clear spiral galaxies that were misclassified by the algorithm. Figure~\ref{kum_elliptical_SDSS} shows galaxies classified as spiral in the \citep{kum16} catalog.

The comparison between the shallow learning and deep neural network shows that while the deep neural network leans towards elliptical galaxies, the shallow learning algorithm is more sensitive to spiral galaxies. The shallow learning algorithm used in \citep{kum16} is described in detail in \citep{orlov2008wnd,shamir2008wndchrm,shamir2010impressionism}. In summary, it computed 2883 numerical image content descriptors from each galaxy image. These image features include edges, textures, fractals, polynomial decomposition, statistical distribution of pixel intensities, and more to provide a comprehensive numerical representation of the image. These features are filtered for the most informative features, weighted for their informativeness, and then classified using an instance-based classifier. Instance-based classifiers have the advantage of handling effectively rare instances, imbalanced classes, and variations inside the classes \citep{zhang2017krnn,li2011improving,mullick2018adaptive}. Since the variability in the class of spiral galaxies is higher than the variability in the class elliptical galaxies, it is possible that the instance-based classifier used in \citep{kum16} can be more accurate in the identification of spiral galaxies.

Experiments done by \cite{walmsley2019identification} compared the accuracy of deep convolutional neural networks to the shallow learning algorithm used in \citep{kum16} for the purpose of automatic morphological classification of galaxies. The results showed that the CNN provided better accuracy compared to the older shallow learning algorithm, especially in cases of faint tidal features. That can explain some of the missclassified galaxies shown in Figure~\ref{kum_spiral}, in which the arms are visible, but are relatively dim. However, the experiments also showed that the shallow learning algorithm used in \citep{kum16} was better able to handle the more complex cases, in which the CNNs struggled to make clear classification \citep{walmsley2019identification}. Since a collection of spiral galaxies is more likely to contain more rare objects, and since the variability among spiral galaxies is higher, an instance-based classifier such as the one used in \citep{kum16} can be more effective in the identification of spiral galaxies compared to elliptical galaxies.





\begin{table}[htbp]
\caption{Examples of images that were misclassified by the model.}
\begin{center}
\begin{tabular}{|c|c||c|c|}
\hline
Misclassified & Confidence & Misclassified  & Confidence \\
as spiral      &                  & as Elliptical     &                 \\
\hline
\includegraphics[scale=0.5]{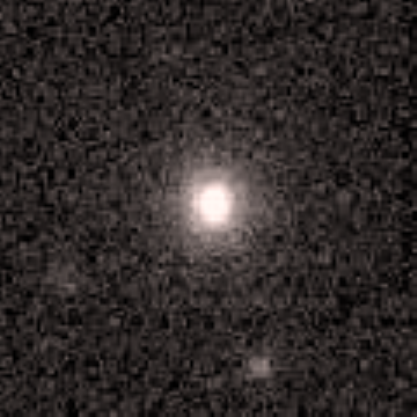} & 0.5181 &
\includegraphics[scale=0.5]{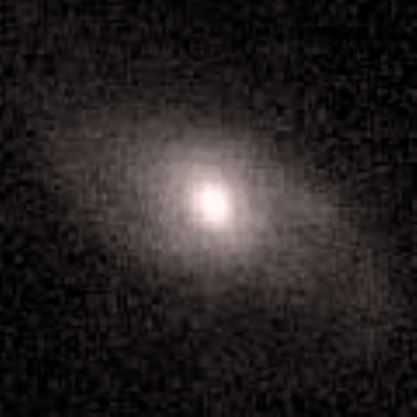} & 0.6017 \\
\hline
\includegraphics[scale=0.5]{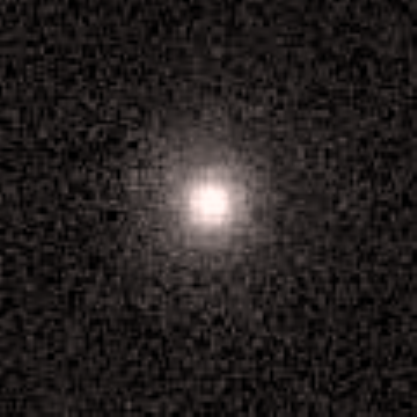} & 0.5614 &
\includegraphics[scale=0.5]{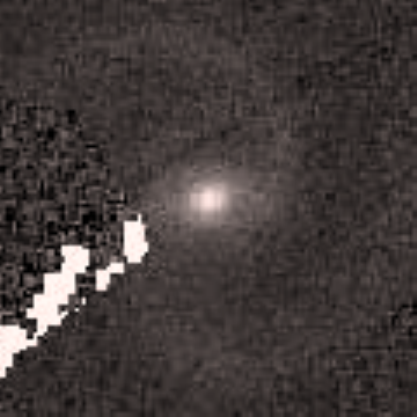}  & 0.6093 \\
\hline
\includegraphics[scale=0.5]{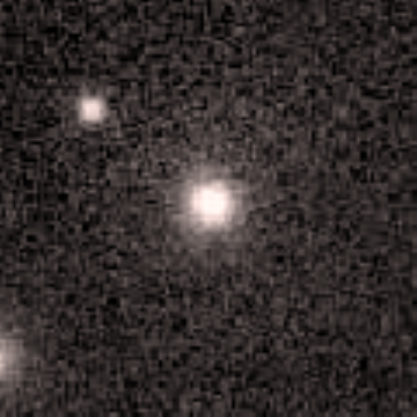} & 0.5940 &
\includegraphics[scale=0.5]{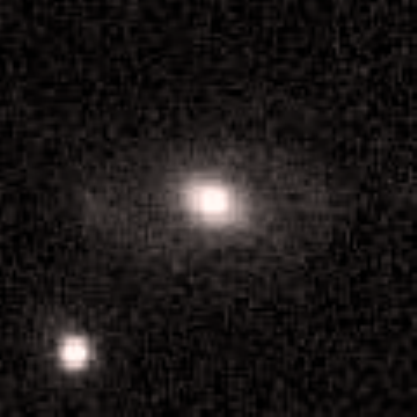} & 0.6455 \\
\hline
\includegraphics[scale=0.5]{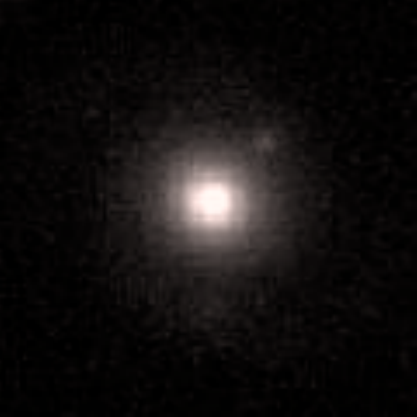} & 0.7677 &
\includegraphics[scale=0.5]{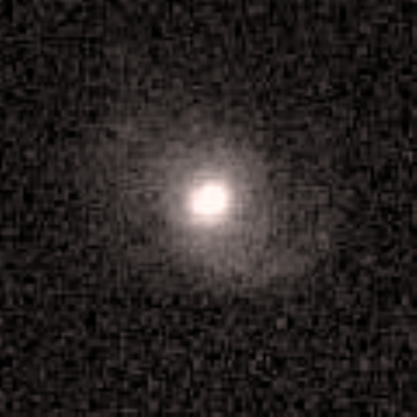} & 0.7493 \\
\hline
\includegraphics[scale=0.5]{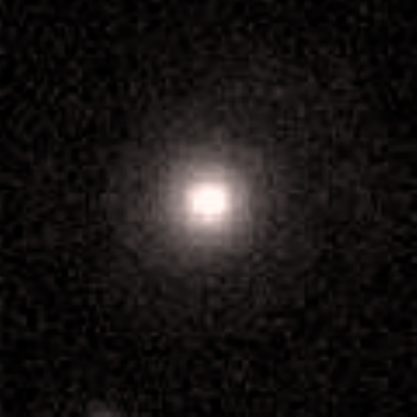} & 0.9114 &
\includegraphics[scale=0.5]{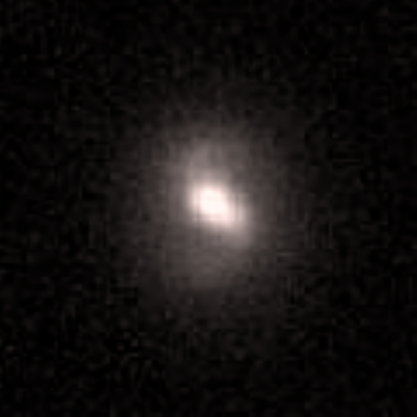} & 0.7647 \\
\hline
\end{tabular}
\label{misclassified_images}
\end{center}
\end{table}


\begin{table}[htbp]
\caption{Examples of images that were classified correctly by the model.}
\begin{center}
\begin{tabular}{|c|c||c|c|}
\hline
Classified & Confidence & Classified  & Confidence \\
as spiral  &                  & Elliptical    &                   \\

\hline
\includegraphics[scale=0.5]{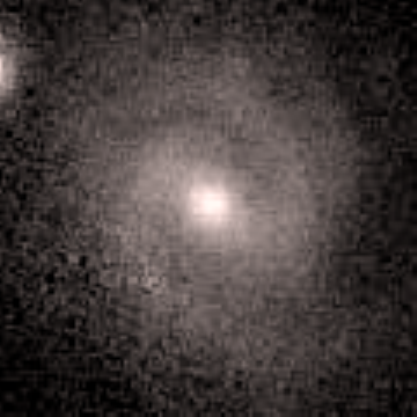}  & 0.9999 &
\includegraphics[scale=0.5]{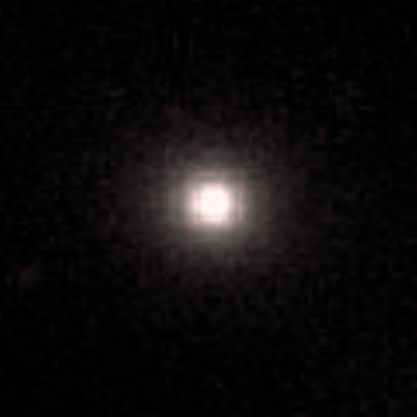} & 0.9999 \\
\hline
\includegraphics[scale=0.5]{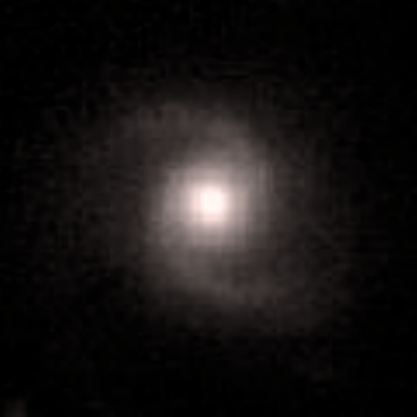}  & 0.9988 &
\includegraphics[scale=0.5]{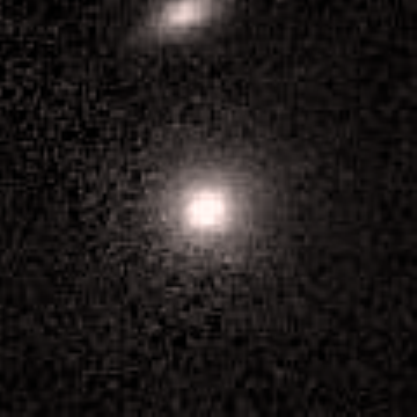} & 0.8799 \\
\hline
\includegraphics[scale=0.5]{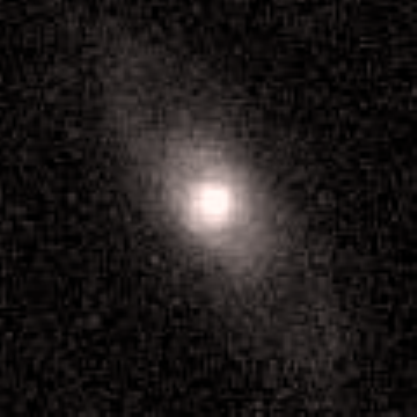} & 0.9718 &
\includegraphics[scale=0.5]{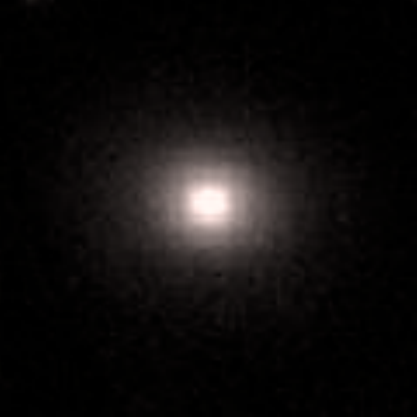} & 0.7568 \\
\hline
\includegraphics[scale=0.5]{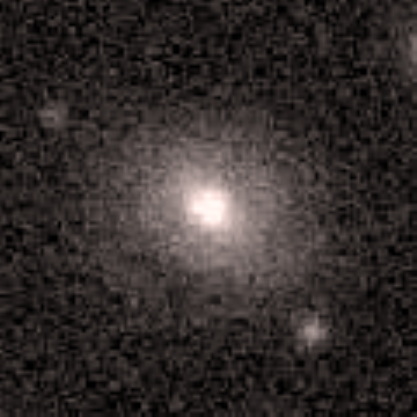} & 0.7446 &
\includegraphics[scale=0.5]{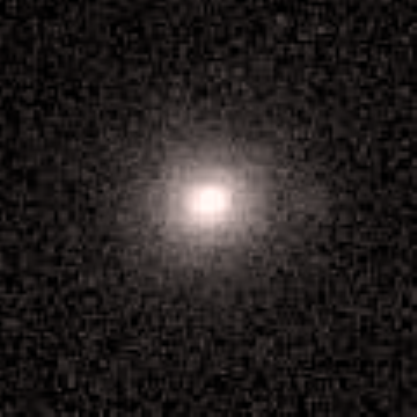} & 0.6780 \\
\hline
\includegraphics[scale=0.5]{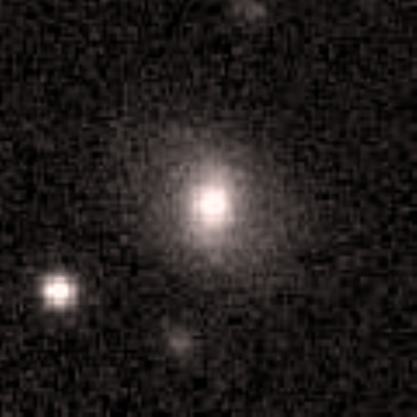} & 0.5163 &
\includegraphics[scale=0.5]{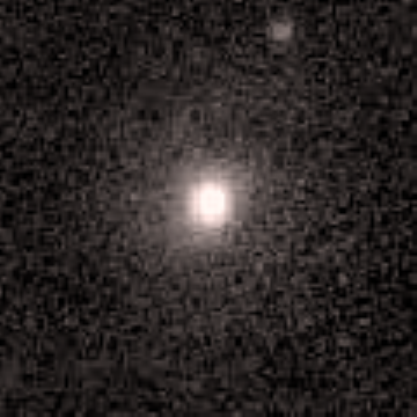} & 0.5637 \\
\hline
\end{tabular}
\label{correct_images}
\end{center}
\end{table}


\section{Conclusions}
\label{conclusions}

While digital sky surveys are capable of collecting and generating extremely large databases, one of the obstacles in fully utilizing these data is the automatic  analysis. Image data, and in particular images of extended objects, are more challenging to analyze due to the complex nature of the image data. Here we created a catalog of Pan-STARRS galaxies classified by their broad morphology into elliptical and spiral galaxies. The likelihood of the annotations provided by the SoftMax layer allows the selection of the objects such that a more consistent catalog by sacrificing some of the galaxies that their classification is less certain. The catalog is available in the form of a CSV file at \url{https://figshare.com/articles/PanSTARRS_DR1_Broad_Morphology_Catalog/12081144}. The classification accuracy is favorably comparable to the $\sim$89\% classification accuracy achieved when using the photometric features provided by the Pan-STARRS photometric pipeline \citep{baldeschi2020star}.

As space-based missions such as Euclid and ground-based missions such as the Rubin Observatory are expected to generate high volumes of astronomical image data, computational methods that can label and organize real-world astronomical images are expected to become increasingly pivotal in astronomy research. Such methods can provide usable data products, and are expected to become important for the purpose of fully utilizing the power of these missions. While convolutional neural networks have demonstrated their ability to classify galaxies by their morphology, a practical solution needs to handle noise, bad data, and inconsistencies that are typical to large real-world datasets. As shown in this paper, the deep neural network is not sufficient to provide clean data products. Instead, a combination of several algorithms that complete a full data analysis pipeline was needed. With the increasing robustness of such systems, it is also expected that protocols that combine multiple neural networks and filtering algorithms will be used to provide detailed morphological information. That information will become part of future data releases of digital sky survey.

The processing was done by first downloading the galaxy images to another server, and the analysis of the data was done on the remote server. The reason for using that practice is because the data analysis is based on solutions designed specifically for the task of galaxy annotation, and not on ``standard'' tasks provided by common services such as CasJobs \citep{li2008casjobs}. Although the smaller JPG images were used, downloading all images still required a substantial amount of time. Using the more informative FITS images would have increased the required time to download the data by an order of magnitude, and analyzing data of much larger digital sky surveys such as the Rubin Observatory will become impractical using this practice. Therefore, future surveys might provide users not merely with certain specific pre-designed tasks, but might also allow processing time for user-designed programs to access the raw data without the need to download it to third-party servers.

\acknowledgments

We would like to thank the anonymous reviewer for the comments that helped to improve the paper. This research was funded by NSF grant AST-1903823. This publication uses data generated via the Zooniverse.org platform, development of which is funded by generous support, including a Global Impact Award from Google, and by a grant from the Alfred P. Sloan Foundation. The Pan-STARRS1 Surveys (PS1) and the PS1 public science archive have been made possible through contributions by the Institute for Astronomy, the University of Hawaii, the Pan-STARRS Project Office, the Max-Planck Society and its participating institutes, the Max Planck Institute for Astronomy, Heidelberg and the Max Planck Institute for Extraterrestrial Physics, Garching, The Johns Hopkins University, Durham University, the University of Edinburgh, the Queen's University Belfast, the Harvard-Smithsonian Center for Astrophysics, the Las Cumbres Observatory Global Telescope Network Incorporated, the National Central University of Taiwan, the Space Telescope Science Institute, the National Aeronautics and Space Administration under Grant No. NNX08AR22G issued through the Planetary Science Division of the NASA Science Mission Directorate, the National Science Foundation Grant No. AST-1238877, the University of Maryland, Eotvos Lorand University (ELTE), the Los Alamos National Laboratory, and the Gordon and Betty Moore Foundation. 

%








\bibliographystyle{apalike}

\bibliography{main}

\end{document}